# Flood zones detection using a runoff model built on Hexagonal shape based cellular automata

Souhaib DOUASS[#1], M'hamed AIT KBIR[*2]

[#]*LIST Laboratory, STI Center, Abdelmalek Essaâdi University*
*UAE, Tangier, Morocco*

***Abstract:*** *This article presents a 3D geographic information systems (GIS) modeling and simulation of water flow in a landscape defined by a digital terrain model, provided by some available geolocation APIs. The proposed approach uses a cellular automata based algorithm to calculate water flow dynamic. The methodology was tested on a case study area of 27kmx19km located in Tangier, north of Morocco. In fact, we aim to detect flood zones in order to prevent problems related to space occupation in urban and rural regions. Some indices can be deduced from the stream shape using Cellular Automata (CA) based approach that can reduce the complexity related to space structures with multiple changes. A spatiotemporal simulation of the runoff process is provided using 3D visualization that we can pair with geographical information system tools (GIS).*

*The 3D GIS modeling approach that was developed for the analyses of flood zones detection using a runoff model based on cellular automata was comprised of three main steps: Input (collection of data), calculation (CA tool) and visualization (3D simulation).*

## I. INTRODUCTION

The simulation based on geographic information systems (GISs) and spatial databases, have progressed but exhibit several characteristics of complexity. Modeling the dynamics of water flow may be a challenging task without the use of tools that are well suited for modeling complex systems. Does not yet allow a better understanding of natural and human phenomena and does not lead to the resolution of urban and rural problems. However, most GISs have a limited capability for handling the temporal component of spatial data. Dynamic approaches, such Cellular Automata and complexity theory, are becoming increasingly used to model the dynamics of spatial complexity of urban systems because of their simple structure, and their capabilities for generating complex patterns that from the local interaction of simple components of the system [1].

The study of the hydrological behavior of the watershed assume the availability of some data matrix (altitude and land parameters) and associated tables including geographical information [2]. These limits appear particularly in spatiotemporal processes that are essentially continuous and dynamic. Among these processes, we can cite global warming, flooding, erosion and urban development. In this article, we propose a 3D modeling approach at a fine spatiotemporal resolution using Cellular Automata simulation tool that we pair with a GIS. We will use a discretization of space with cells, using a surface approximatively equal to $14 \ m^2$ . We propose to simulate the flow of water on the surface of the Tangier region, north of Morocco, with a geographical area of a dimension (27kmx19km) and we simulate the rain with an average of 80mm accumulated during 12h and 24h.

In the last two decades, the most important risks, in terms of potential human and economic impacts, are the risks of flooding. Among the significant damage caused by the floods, we can cite the impact on the highway infrastructure. For example, the two frequently used highway sections: Tangier-Port and Tetouan-Fnideq, located at the level of the unstable Rif massif, suffered damage characterized mainly by embankment gullies, degradations of some sanitation work and localized landslides of excavation embankment materials [3].

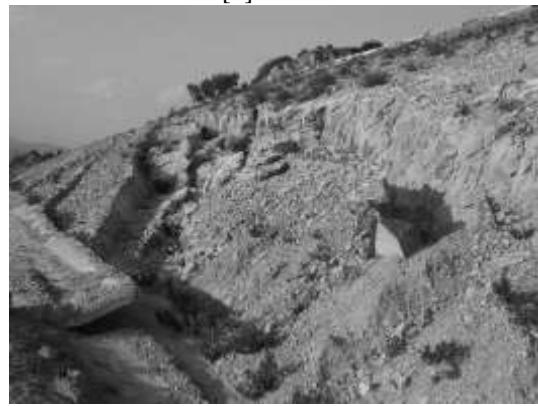

**Fig 1 : Unhooking of materials from an excavation slope (Tetouan–Fnideq highway section)**

We also cite, the strong floods which marked the period 2008-2010, how were at the origin of punctual deposits on the roadway of the two cited sections, because of the obturation of the works of remediation with bed load material from the cut slopes.

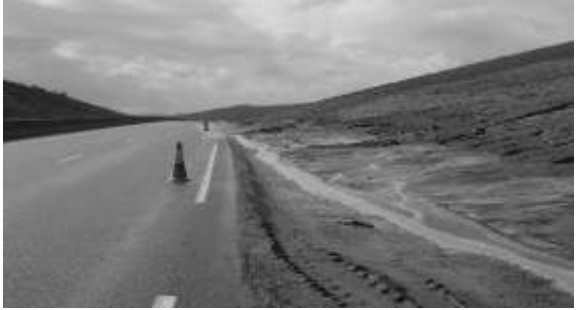

**Fig 2 : Tangier-Port highway (mileage point 40 towards the Port)**

The consequences engendered on the traffic, at the level of Tangier in 2008, was estimated to 2 mittions MAD, Moroccan Dirham, (more than 200 thousand $) of damage caused by the overflow of Moghogha river. That caused circulation stop between some cities for more than 2 days. In 2010, damages are estimated to 15 million MAD, due to the exceptional rainfall and the erodible nature of the land of the Rif (mainly pellitic), favoring the filling of the sanitation works, so that it produced a stop of the circulations between Tangier city and Tangier Med port for 2 days [3].

On Monday November 21, 2016, the water level reached 72 millimeters, according to the direction of National Meteorology. A few hours of precipitation were enough for Tangier to be submerged by the waters. Despite recent investments in infrastructure upgrades, the same scenario is repeated every time.

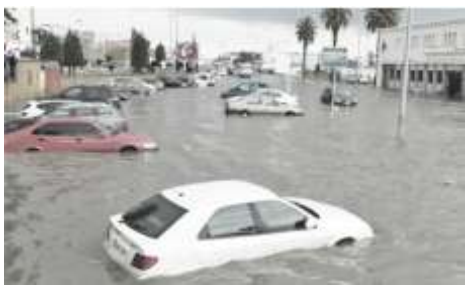

**Fig 3 : City's roads which could not absorb the volume of precipitation [4]**

Morocco has a national flood control plan that covers more than 390 threatened sites, but it is not fully prepared for it. Prevention remains little present in public policies and in municipal development plans. Hence the urgency of actions to raise awareness of the need to act, in particular to develop solutions based on new information technologies.

In a previous paper [5], we worked with Cellular Automata with regular rectangular grids. The same digital terrain model is used in the present paper, we are looking forward to improve the system and get more relevant results.

## II. DIGITAL TERRAIN MODEL

In representing the terrain surface, the digital terrain model (DTM) is one of the most important tools.

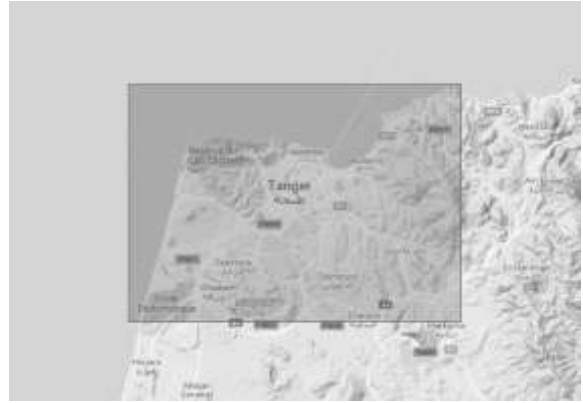

**Fig 4 : Zone area under study**

In our works, we used Google map API, to get the geographic coordinates of a given area, each position is defined by longitude and altitude.

We have divided this rectangular area into a two-dimensional grid, each grid element contains a position in longitude and latitude, then through the Google Maps API Elevation Service [6] we obtained, the height in meters of each position.

The distance between two points is the length of the shortest path between them. This shortest path is called a geodesic. On a sphere all geodesics are segments of a great circle. To compute this distance, we use functions of Google Geometry Library [7].

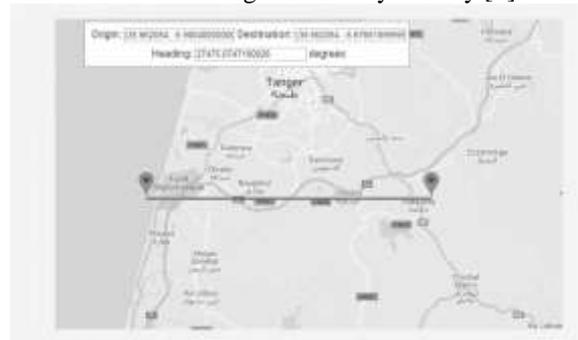

**Fig 5 : Length in the area, defined by two geographical positions**

To normalize the 3D terrain model, we set the model width and height on a scale unit [8]. The real dimensions of this zone is 27km x 19km.

In the following figure we visualize the 3D terrain model using a WebGL application.

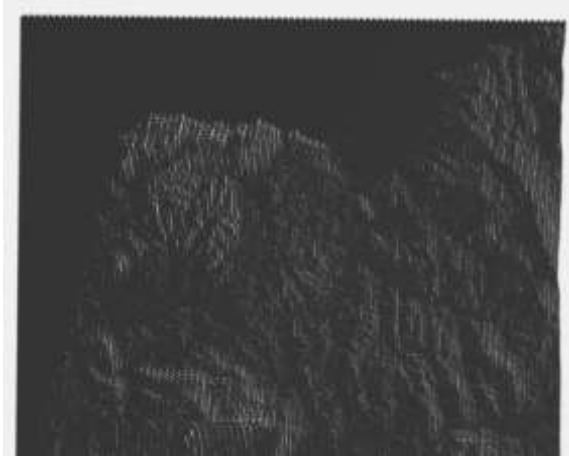

**Fig 6 : The 3D digital of terrain model**

The following graph shows elevation variation expressed in meters of 100 points taken from the diagonal line of the considered rectangular terrain, with respect to the South-Est /North-West orientation. We note that it is an area with chaotic relief which favors strong flows and the accumulation of water in low altitude areas in record time.

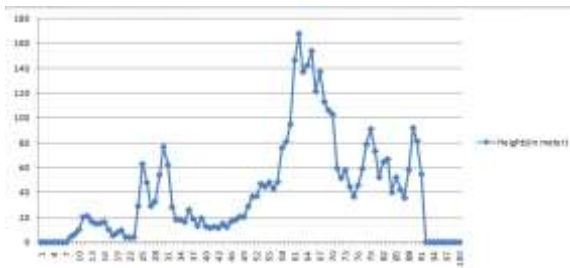

**Fig 7 : Variation of elevation in meter**

### III. WEBGL TECHNOLOGY

In 2009, Khronos established the WebGL working group and then started the standardization process of WebGL based on OpenGL ES 2.0, releasing the first version of WebGL in 2011.

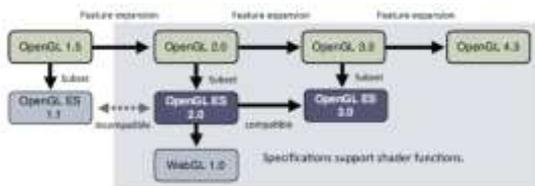

**Fig 8 : Relationship among OpenGL, OpenGL ES 1.1/2.0/3.0 and WebGL.**

WebGL runs on the GPU on computer, there are two functions are called, a vertex shader (it's compute vertex positions) and a fragment shader (its computes a color for each pixel), they are written in GLSL (GL Shader Language). Web pages using WebGL are created by using three languages: HTML5 (as a Hypertext Markup Language), JavaScript and GLSL ES [9].

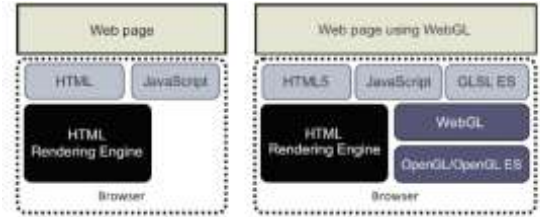

**Fig 9 : The software architecture of dynamic web pages (left) and web pages using WebGL**

### IV. HEXAGONAL CELLULAR AUTOMATA

#### A. Cellular Automata basics

Cellular automata was used to solve several problems in different fields, for example, we can cite the flow simulations on the earth surface, optimization computation process [10], urban growth simulation [11] and modeling of erosive runoff [11].

The important principles of CA are: Spatial structure, local interaction and cells state variation over time.

In this we need to realize the water flow simulation system with an interactive model using CA. For this, we define a grid of cells, each cell corresponds to a small area, with a state, water level, that can change depending on its own state and its neighbors state.

CA models are based on the use of several primary components. These components are essentially centered on cells that are arranged in a regular grid with transition rules determining changes in cells properties. These components affect the status of each individual cell in a network at each time step.

Therefore, the three fundamental features of a CA are:

Uniformity: all cell states are updated by the same set of rules.

Synchronicity: all cell states are updated simultaneously

Locality: rules are local at each cell

In the flowing figure, we have an example of a two-dimensional configuration with each cell taking one of two state values, blue or not (having water or not), according to a particular local transition function [12].

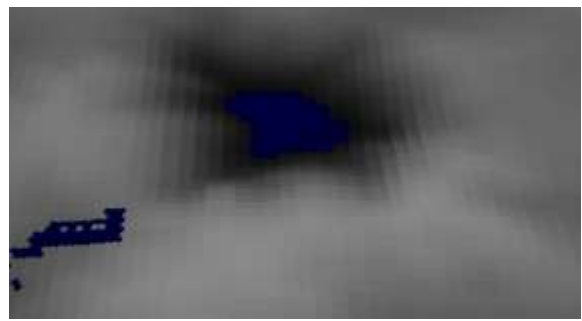

.-a-

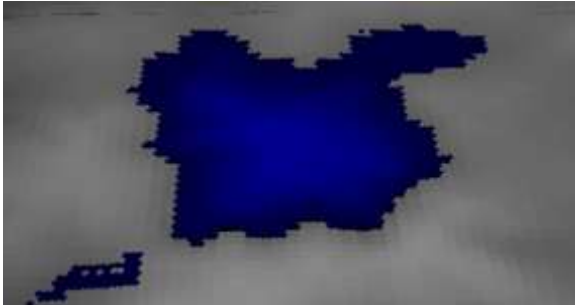

-b-

**Fig 10 : A simple two-dimensional grid, each cell (blue cells) taking a value that correspond to the water level**

The analysis of the behavior of a hydrological system is most often carried out by studying the reaction of the terrain under study against precipitations. This reaction is measured by observing the amount of water flowing to the system outlets.

### B. Neighbor shape

A cellular world has neighbors in each side, self-organization in a complex system, raises in local interaction between cells. This local interaction is a second important principal of Cellular Automata model that allows them to simulate real dynamic systems.

In the figures bellow we can see different type of neighbors often used in cellular automata model [13].

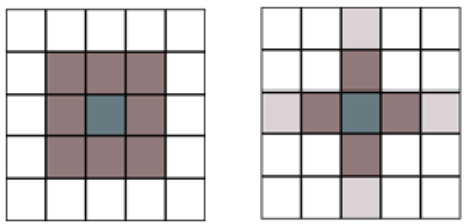

**Fig 11 : 8, 4 neighbors for the blue cell.**

We can present a cellular world by a hexagonal grid, in this case, each cell has six nearest neighbors [14]. As shown in the following figure.

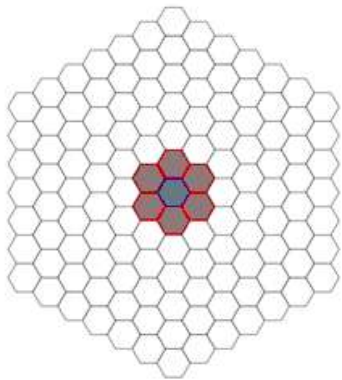

**Fig 12 : 6 neighbors for the blue cell in the hexagonal grid**

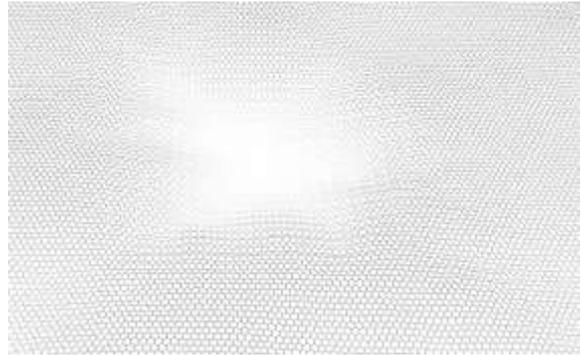

**Fig 13 : The six nearest neighbours of a hexagonal cells of a simple landscape.**

The importance of a hexagonal grid against a square grid is that the distance between the center of each and every pair of adjacent hex cells is the same. By comparison, in a square grid map, the distance from the center of each square cell to the center of the four diagonal adjacent cells, with whom it shares a single corner, is greater than the distance to the four adjacent cells, with whom it shares an edge. The other advantage is the fact that neighboring cells always share edges; there are no two cells with a single common point like in a square grid.

### C. Hexagonal grid

There is one obvious way to tune in a square grid. But with hexagons, there are multiple approaches, like cube coordinates algorithms, offset coordinates, Axial coordinates or Doubled coordinates [15].

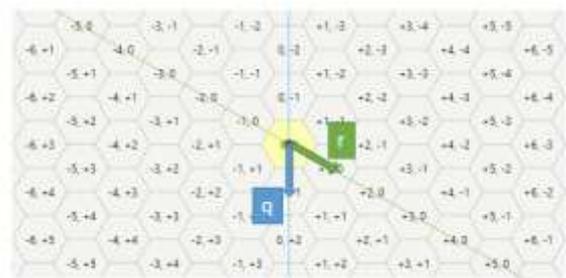

**Fig 14 : Hexagon metrics**

To compute hexagon face centers, there's an q basis vector (x=$\sqrt{3}$, y=0) and r basis vector (x=$\frac{\sqrt{3}}{2}$, y=3/2), as illustrated in the Figure 14. Going from hexagonal coordinates to world coordinates is a matrix multiply:

$$\begin{pmatrix} x \\ y \end{pmatrix} = size * \begin{bmatrix} \frac{3}{2} & 0 \\ \frac{\sqrt{3}}{2} & \sqrt{3} \end{bmatrix} * \begin{pmatrix} q \\ r \end{pmatrix}$$

Expanded out, that's:

$$x = size * (\sqrt{3} * q + \frac{\sqrt{3}}{2} * r)$$
$$y = size * (\frac{3}{2} * r)$$

Going from hexagonal coordinates (q, r) to world coordinates (x, y) was done by matrix multiplication, we can solve the equations for (q, r):

$$\begin{pmatrix} q \\ r \end{pmatrix} = \begin{bmatrix} \dfrac{2}{3} & 0 \\ -\dfrac{1}{3} & \dfrac{\sqrt{3}}{2} \end{bmatrix} * \begin{pmatrix} x \\ y \end{pmatrix} / size$$

The resulting code is:

$$q = \left(\frac{2}{3} * x\right) / size$$
$$r = \left(-\frac{3}{2} * x + \frac{\sqrt{3}}{2} * y\right) / size$$

This calculation gives a fractional axial coordinates (floats) for q and r. then we convert(round) the fractional axial coordinates into integer axial hex coordinates.

### D. Water distribution rule

Several partitions variants were investigated, leading to different neighborhood geometries as shown in previous part. In the method presented in this paper for pattern generation, each cell is connected to its six closest neighbors in a hexagonal grid. The basic partition adopted in the present study as shown in figure 15.

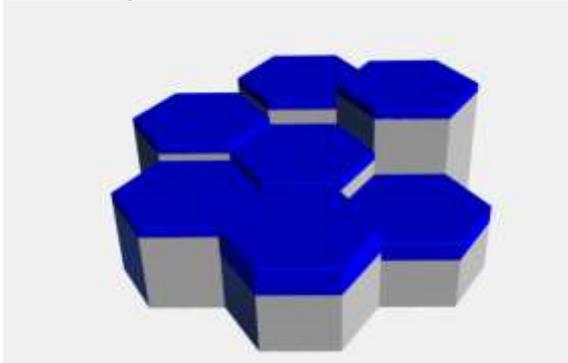

**Fig 15 : Basic partition adopted in the present study**

By means of the basic water distribution rule, the water is distributed inside each basic partition according to the following procedure [16], initially adopted to square grids:

1. Order the cells index according to the terrain height (see Figure 16), that is:

$$h_1 \le h_2 \le \cdots \le h_7$$

2. Let $w_i^{old}$ be the current water level of the i-cell.
3. Calculate the equilibrium surface water height, H, as the height the water would reach, provided that the total water volume contained in the partition drains down to the lowest possible locations, that is the equation:

$$H = \frac{W + \sum_{i=1}^{k} h_i}{k}$$

4. Where W is the water volume contained in the partition divided by the unit-cell area that can be calculated as:

$$W = \sum_{i=1}^{7} (w_i^{old} - h_i)$$

5. And k is the number of cells that remains wet after the water drains down, which is the largest cell index satisfying:

$$W \ge \sum_{i=1}^{7} (h_k - h_i)$$

6. Calculate the drained water level $w_i^{drain}$, that would have the cells if the water volume contained in the partition drains down to the lowest locations:

$$w_i^{drain} = H - h_i \qquad if \ i \le k$$
$$w_i^{drain} = 0 \qquad if \ i > k$$

## V. APPLICATION AND RESULTS

We simulate the rain with an average of 80mm, accumulated during 3 days in our geographical area. Each cell has a bounding box of 4 meters by 4 meters in the 2D geographical map, its meaning that the width of a hexagon W=4 meters and according to these 2 formulas, we calculate the area of a regular hexagon $A = 13.85 \ m^2$ and size of an edge S = 4.61 meters :

$$W = \sqrt{3} * S$$
$$A = \frac{3\sqrt{3}}{2} * S^2$$

The water level contained in each cell is modified by $1.11 \ 10^{-3} \ m^3/$ hour. We suppose also that cells updating and water distribution is done each 10s (calculation step). After each step all cells are updated in a random order.

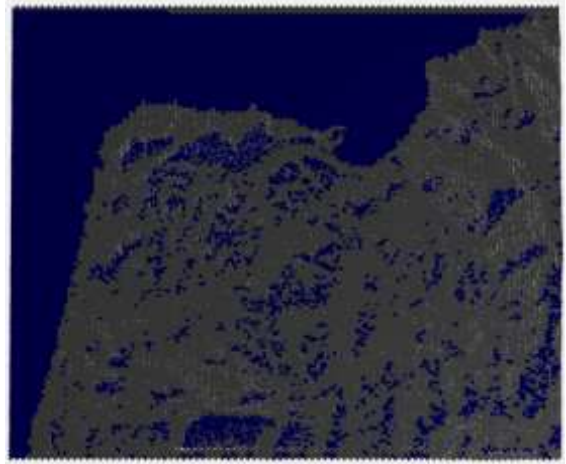

**Fig 16 : Simulation of water flow after 12 hours of rain**

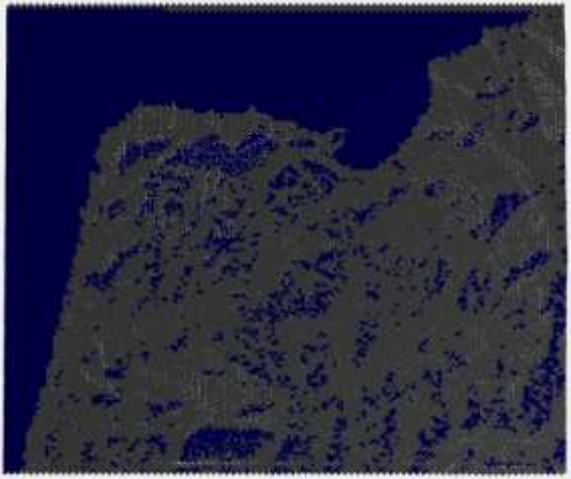

**Fig 17 : Simulation of water flow after 24 hours of rain**

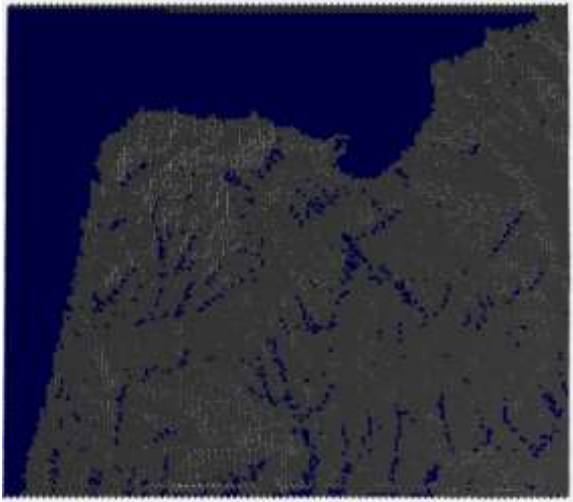

**Fig 18 : Simulation of water flow after 12 hours of rain and 4 hours of equilibrate**

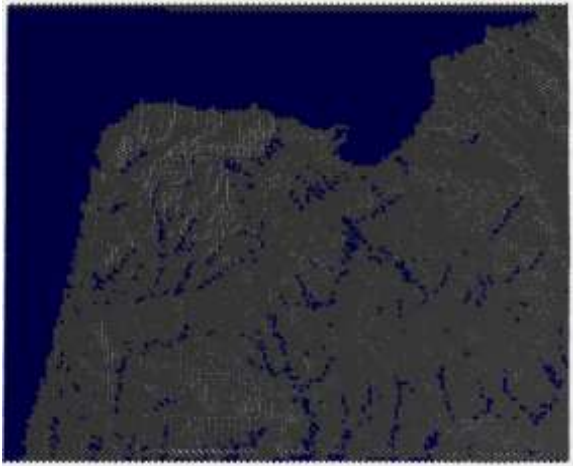

**Fig 19 : Simulation of water flow after 24 hours of rain and 4 hours of equilibrate**

According to these simulation results, we find that the areas at high risk are:

- The region of "Moghora" with a maximum water depth of (2 -4.8 m).
- The region of "Aouama" with a maximum water depth of (2 -3 m).

We are selecting for water level some analysis and performing various calculations:

- Count of the cells: 21033
- Total sum of the water values: 597.884 meters
- Sample standard deviation of the water values: 0.173379 meters
- Mean of the water values: 0.028426 meters
- Minimum of the water values: 0.000000 meters
- Maximum of the water values: 4.818000 meters

These results fit well with reality, and correspond, among other things, to rivers that cross a part of the city. Their floods threaten the neighborhoods they pass through, which requires hydraulic facilities to protect against floods. It can also be a question of building scuppers of a suitable height, or the deviation and the search for alternative solutions for the networks plans, of any type, in progress.

## VI. CONCLUSIONS

This work demonstrates the feasibility of using Cellular Automata based on a regular hexagonal area, coupled with a GIS for water flow and consequently flood simulation.

The CA in this article is the initial, work to design a more complex natural process, which could be used to model different kinds of problems, such as air pollution.

In the perspective for this work, we are planning to work with a Cellular Automata based on a hexagonal grid, and take into account new parameters when elaborating transition rules like vegetation cover type, soil characteristics, environmental constraints and rainfall irregularity.